\documentclass[aps,pre,floats,preprint,superscriptaddress]{revtex4}

\usepackage{graphicx,epsfig}
\usepackage{rotate}
\usepackage{amsmath}
\usepackage{mathrsfs}
\usepackage{times}
\usepackage{graphics,dcolumn,bm,fleqn,epic,eepic,float}
\usepackage{amssymb,amsmath,multirow,rotate,color}

\newcommand{\be}{\begin{equation}}
\newcommand{\ee}{\end{equation}}

\begin{document}

\title{Cluster structure of functional networks \\ estimated from high-resolution EEG data}

\author{Roberta Sinatra\footnote{To whom corrispondence should be addressed. E-mail: robertasinatra@gmail.com, rsinatra@ssc.unict.it}}\thanks{These authors have contributed equally to the present paper (co-first authors).}
\affiliation{Scuola Superiore di Catania, Catania, Italy}

\author{Fabrizio De Vico Fallani}\thanks{These authors have contributed equally to the present paper (co-first authors).}
\affiliation{Interdepartmental Research Centre for Models and Information Analysis in Biomedical Systems, University ``La Sapienza'', Rome, Italy}
\affiliation{IRCCS ``Fondazione Santa Lucia'', Rome, Italy}

\author{Laura Astolfi}
\affiliation{IRCCS ``Fondazione Santa Lucia'', Rome, Italy}
\affiliation{Department of Human Physiology and Pharmacology, University ``La Sapienza'', Rome, Italy}

\author{Fabio Babiloni}
\affiliation{IRCCS ``Fondazione Santa Lucia'', Rome, Italy}
\affiliation{Department of Human Physiology and Pharmacology, University ``La Sapienza'', Rome, Italy}

\author{Febo Cincotti}
\affiliation{IRCCS ``Fondazione Santa Lucia'', Rome, Italy}

\author{Vito Latora}
\affiliation{Department of Physics and Astronomy, University of Catania, and INFN Sezione Catania, Italy}

\author{Donatella Mattia}
\affiliation{IRCCS ``Fondazione Santa Lucia'', Rome, Italy}

\date{\today}

\begin{abstract}
We study the topological properties of functional connectivity patterns among cortical areas in the frequency domain. The cortical networks were estimated from high-resolution EEG recordings in a group of spinal cord injured patients and in a group of healthy subjects, during the preparation of a limb movement. We first evaluate global and local efficiency, as indicators of the structural connectivity respectively at a global and local scale. Then, we use the Markov Clustering method to analyse the division of the network into community structures. The results indicate large differences between the injured patients and the healthy subjects. In particular, the networks of spinal cord injured patient exhibited a higher density of efficient clusters. In the Alpha (7-12 Hz) frequency band, the two observed largest communities were mainly composed by the cingulate motor areas with the supplementary motor areas, and by the pre-motor areas with the right primary motor area of the foot. This functional separation strengthens the hypothesis of a compensative mechanism due to the partial alteration in the primary motor areas because of the effects of the spinal cord injury. 

\end{abstract}

\maketitle

\section{Introduction}
Over the last years, there has been an increasingly large interest in finding significant features from human brain networks. In particular, the concept of functional connectivity plays a central role to understand the organized behaviour of anatomical regions in the brain during their activity. This  organization is thought to be based on the interaction between different and differently specialized cortical sites. Indeed, several methods have been proposed and discussed in the literature, with the aim of estimating the functional relationships among the physiological signals [David \emph{et al.}, 2004; Lee \emph{et al.}, 2003] obtained from different neuro-imaging devices such as the functional Magnetic Resonance Imaging (fMRI) scanner, electroencephalography (EEG) and magnetoencephalography (MEG) apparatus [Horwitz, 2003]. Recently, a multivariate spectral technique called Directed Transfer Function (DTF) has been proposed [Kaminski \emph{et al.}, 2001] to determine directional influences between any given pair of channels in a multivariate data set. This estimator is able to characterize at the same time direction and spectral properties of the brain signals, requiring only one multivariate autoregressive (MVAR) model to be estimated from all the EEG channel recordings. The DTF index has been demonstrated [Kaminski \emph{et al.}, 2001] to rely on the key concept of Granger causality between time series -an observed time series $x(n)$ causes another series $y(n)$ if the knowledge of $x(n)$’s past significantly improves prediction of $y(n)$- [Granger, 1969]. However, the extraction of salient characteristics from brain connectivity patterns is an open challenging topic, since often the estimated cerebral networks have a relative large size and complex structure. Consequently, there is a wide interest in the development of mathematical tools that could describe in a concise way the structure of the estimated cerebral networks [Tononi \emph{et al.}, 1994; Stam, 2004; Salvador, 2005; Sporns, 2002]. 

Functional connectivity networks estimated from EEG or magnetoencephalographic (MEG) recordings can be analyzed with tools that have been already proposed for the treatments of complex networks as graphs [Strogatz, 2001; Wang \& Chen, 2003; Sporns \emph{et al.}, 2004; Stam \emph{et al.}, 2006]. Such an approach can be useful, since the use of mathematical measures summarizing graph properties allows for the generation and the testing of particular hypothesis on the physiologic nature of the functional networks estimated from high-resolution EEG recordings. However, first results have been obtained for a set of anatomical brain networks [Strogatz, 2001; Sporns \emph{et al.}, 2002]. In these studies, the authors have employed two characteristic measures, the \emph{average shortest path L} and \emph{the clustering index C}, to extract respectively the global and local properties of the network structure [Watts and Strogatz 1998]. They have found that anatomical brain networks exhibit many local connections (i.e. a high $C$) and a shortest separation distance between two randomly chosen nodes (i.e. a low $L$). Hence, anatomical brain networks have been designated as “small-world” in analogy with the concept of the small-world phenomenon observed more than 30 years ago in social systems [Milgram 1967]. 

Many types of functional brain networks have been analyzed in a similar way. Several studies based on different imaging techniques like fMRI [Salvador \emph{et al.}, 2005; Eguiluz \emph{et al.}, 2005; Achard \& Bullmore, 2007], MEG [Stam \emph{et al.}, 2006; Bassett \emph{et al.}, 2006; Bartolomei \emph{et al.}, 2006] and EEG [Micheloyannis \emph{et al.}, 2006; Stam \emph{et al.}, 2007] have shown that the estimated functional networks can indeed exhibit the small-world property. In the functional brain connectivity context, these properties have been demonstrated to reflect an optimal architecture for the information processing and propagation among the involved cerebral structures [Lago-Fernandez \emph{et al.}, 2000; Sporns \emph{et al.}, 2000]. 
In particular, a high \emph{clustering index C} is an indication of the presence in the network of a large number of triangles. However, this index alone does not return detailed information on the presence of larger connected clusters of nodes. This fact makes up a real obstacle in the analysis of the network properties especially in the field of the Neuroscience where the correlated behaviour of different cortical regions plays a fundamental role in the correct understanding of cerebral systems. Methods to detect the community structures in a graph, i.e. tightly connected group of nodes, are now available in the market [Harary \& Palmer, 1973]. Communities (or clusters or modules) are groups of vertices that probably share common properties and/or play similar roles within the graph [Boccaletti \emph{et al.}, 2006]. Hence, communities may correspond to groups of pages of the World Wide Web dealing with related topics [Flake \emph{et al.}, 2002], to functional modules such as cycles and pathways in metabolic networks [Guimerà \& Amaral, 2005; Palla \emph{et al.}, 2005], to groups of affine individuals in social networks [Girvan and Newman, 2002; Lusseau and Newman, 2004], to compartments in food webs [Pimm, 1979; Krause \emph{et al.}, 2003], and so on.
Finding the communities within a cerebral network allows identifying the hierarchy of functional connections within a complex architecture. This opportunity would represent an interesting way to improve the basic understanding of the brain functioning. Indeed, some cortical regions are supposed to share a large number of functional relationships during the performance of several motor and cognitive tasks. This characteristic leads to the formation of highly connected clusters within the brain network. These functional groups consist in a certain number of different cerebral areas that are cooperating more intensively in order to complete a task successfully.

In the present paper, we present a study of the structural properties of functional networks estimated from high-resolution EEG signals in a group of spinal cord injured patients during the preparation of a limb movement. In particular, we first investigate some indicators of the connectivity at a global and local scale. Then we analyse the networks by studying their structure in communities, and we compare the results with those obtained from a group of healthy subjects.

\section{Methods}
\subsection{High-resolution EEG recordings in SCI patients and Healthy subjects}
All the experimental subjects participating in the study were recruited by advertisement. Informed consent was obtained in each subject after the explanation of the study, which was approved by the local institutional ethics committee
The spinal cord injured (SCI) group consisted of five patients (age, 22-25 years; two females and three males). Spinal cord injuries were of traumatic aetiology and located at the cervical level (C6 in three cases, C5 and C7 in two cases, respectively); patients had not suffered for a head or brain lesion associated with the trauma leading to the injury.
The control (CTRL) group consisted of five healthy volunteers (age, 26-32 years; five males). They had no personal history of neurological or psychiatric disorder; they were not taking medication, and were not abusing alcohol or illicit drugs. 
For EEG data acquisition, subjects were comfortably seated on a reclining chair, in an electrically shielded, dimly lit room. They were asked to perform a brisk protrusion of their lips (lip pursing) while they were performing (healthy subjects) or attempting (SCI patients) a right foot movement.
The choice of this joint movement was suggested by the possibility to trigger the SCI’s attempt of foot movement. In fact patients were not able to move their limbs; however they could move their lips. By attempting a foot movement associated with a lips protrusion, they provided an evident trigger after the volitional movement activity. This trigger was recorded to synchronize the period of analysis for both the considered populations.
The task was repeated every 6-7 seconds, in a self-paced manner, and the 100 single trials recorded will be used for the estimate of functional connectivity by means of the Directed Transfer Function (DTF, see following paragraph). A 96-channel system (BrainAmp, Brainproducts GmbH, Germany) was used to record EEG and EMG electrical potentials by means of an electrode cap and surface electrodes respectively. The electrode cap was built accordingly to an extension of the 10-20 international system to 64 channels. Structural MRIs of the subject’s head were taken with a Siemens 1.5T Vision Magnetom MR system (Germany).

\subsection{Cortical activity and Functional connectivity Estimation}
 Cortical activity from high resolution EEG recordings was estimated by using realistic head models and cortical surface models with an average of 5.000 dipoles, uniformly disposed. Estimation of the current density strength, for each one of the 5.000 dipoles, was obtained by solving the Linear Inverse problem, according to techniques described in previous papers [Babiloni \emph{et al.}, 2005; Astolfi \emph{et al.}, 2006]. By using the passage through the Tailairach coordinates system, twelve Regions Of Interest (ROIs) were then obtained by segmentation of the Brodmann areas on the accurate cortical model utilized for each subject. The ROIs considered for the left (\_L) and right (\_R) hemisphere are: the primary motor areas for foot (MF\_L and MF\_R) and lip movement (ML\_L and ML\_R); the proper supplementary motor area (SM\_L and SM\_R); the standard pre-motor area (6\_L and 6\_R); the cingulated motor area (CM\_L and CM\_R) and the associative Brodmann area 7 (7\_L and 7\_R). For each EEG time point, magnitude of the five thousands dipoles composing the cortical model was estimated by solving the associated Linear Inverse problem [Grave de Peralta \& Gonzalez Andino, 1999]. Then, the average activity of dipoles within each ROI was computed. In order to study the preparation to an intended foot movement, a time segment of 1.5 seconds before the lips pursing was analysed; lips movement was detected by means of an EMG. The resulting cortical waveforms, one for each predefined ROI, were then simultaneously processed for the estimation of functional connectivity by using the Directed Transfer Function. The DTF is a full multivariate spectral measure, used to determine the directed influences between any given pair of signals in a multivariate data set [Kaminski \emph{et al.}, 2001].  In order to be able to compare the results obtained for data entries with different power spectra, the normalized DTF was adopted. It expresses the ratio of influence of element $j$ to element $i$ with respect to the influence of all the other elements on $i$.  Details on the DTF equations in the treatment of EEG signals have been largely described in previous papers [Astolfi \emph{et al.} 2005, Babiloni \emph{et al.} 2005]. In the present study, we selected four frequency bands of interest (Theta 4-7 Hz, Alpha 8-12 Hz, Beta 13-29 Hz and Gamma 30-40 Hz) and we gathered the respective cortical networks by averaging the values within the respective range. In order to consider only the functional links that are not due to chance, we adopted a Montecarlo procedure. In particular, we contrasted each DTF value with a surrogate distribution of one thousand DTF values obtained by shuffling the signals’ samples in the original EEG dataset. Then, we considered a threshold value by computing the 99$^{th}$ percentile of the distribution and we filtered the original DTF values by removing the edges with intensity below the statistical threshold. 

\subsection{Evaluation of global and local efficiency}
A graph is an abstract representation of a network. A graph $G$ consists in a set of vertices -or nodes- $\mathscr{V}$ and a set of edges -or connections- $\mathscr{L}$ indicating the presence of some sort of interaction between the vertices. A graph can be described in terms of the so-called adjacency matrix $A$, a square matrix such that, when a weighted and directed edge exists from the node $i$ to $j$, the corresponding entry of the adjacency matrix is $A_{ij} \neq 0$; otherwise $A_{ij} = 0$. Two measures are frequently used to characterize the local and global structure of unweighted graphs: the average shortest path $L$ and the clustering index $C$ [Watts and Strogatz, 1998; Newman, 2003; Grigorov, 2005]. The former measures the average distance between two nodes, the latter indicates the tendency of the network to form highly connected clusters of nodes. Recently, a more general setup has been proposed to study weighted (also unconnected) networks [Latora \& Marchiori, 2001; Latora \& Marchiori, 2003]. The efficiency $e_{ij}$ in the communication between two nodes $i$ and $j$, is defined as the inverse of the shortest distance between the vertices. Note that in weighted graphs the shortest path is not necessarily the path with the smallest number of edges. In the case the two nodes are not connected, the distance is infinite and $e_{ij} = 0$. The average of all the pair-wise efficiencies $e_{ij}$ is the global-efficiency $E_g$ of the graph $G$:
\begin{equation}
 E_g(G) = \frac{1}{N(N-1)}\sum_{i \neq j \in \mathscr{V}} {\frac{1}{d_{ij}}}
\end{equation}
where $N$is the number of vertices composing the graph. The local properties of the graph can be characterized by evaluating for every vertex $i$ the efficiency of $G_i$, which is the sub-graph induced by the neighbours of the node $i$ [Latora \& Marchiori, 2001]. Thus, we defined the local efficiency $E_l$ of graph $G$ as the average:
\begin{equation}
 E_l(G) = \frac{1}{N}\sum_i {E_g(G_i)}
\end{equation}
Since node $i$ does not belong to subgraph $G_i$, the local efficiency measures how the communication among the first neighbours of $i$ is affected by the removal of $i$ [Latora and Marchiori, 2005; Latora and Marchiori, 2007]. Hence the local efficiency is an indicator of the level of fault-tolerance of the system. Separate ANOVAs were conduced for each of the two variables $E_g$ and $E_l$. Statistical significance was fixed at 0.05, and main factors of the ANOVAs were the ``between'' factor GROUP (with two levels: SCI and CTRL) and the ``within'' factor BAND (with four levels: Theta, Alpha, Beta and Gamma). Greenhouse \& Geisser correction has been used for the protection against the violation of the sphericity assumption in the repeated measure ANOVA. Besides post-hoc analysis with the Duncan's test and significance level at 0.05 has been performed. 


\subsection{Detection of Community Structures}
\label{detection}
In order to detect the community structures, we have implemented the Markov Clustering (MCl) algorithm [Vandongen PhD Thesis, 2000; Enright \emph{et al.}, 2002]. It is one algorithm of a few available which works even with directed graphs and it is based on the properties of the dynamical evolution of random walkers moving on the graph. This approach is also useful since it manages to achieve reliable results when the graph contains self-loops, i.e. edges connecting a node to itself. Since a community is a group of densely connected nodes, a random walker that started in a node of a given community will leave this cluster only after having visited a large number of the community's nodes. Hence, the basic idea implemented in the algorithm is to favour the random motion within nodes of the same community. This is obtained by alternating the application of two operators on the transition matrix of the random walk: the expansion operator and the inflation one. The expansion operator applied on a given matrix returns its square power, while the inflation operator corresponds to the Hadamard power of the same matrix, followed by a scaling. In practice, the algorithm works as follows:
\begin{enumerate}
\item Take the adjacency matrix $A$ and add a self-loop to each node, i.e. set $A_{ii}=1$ for $i=1,2,\dots,N$; 
\item Obtain from $A$ the transition probability matrix $W$, that describes the random motion: $W_{ij}=\frac{A_{ij}}{\sum_k{A_{kj}}}$. Every element $W_{ij}$ expresses the probability to go from $j$ to $i$ in one step. $W$ is a stochastic matrix i.e. a matrix of non-negagtive elements and where the sum of elements of each column is normalized to one: $\sum_{i=1}^{N}{W_{ij}}=1$;
\item Take the square of $W$ (expansion step);
\item Take the r$^{th}$ power ($r>1$) of every element of $W^2$ and normalize each column to one to obtain a new stochastic matrix $W'$: $W'=\frac{[(W^2)_{ij}]^r}{\sum_k{[(W^2)_{kj}]^r}}$ (inflation step);
\item Go back to step 3.
\end{enumerate}

Step 3 corresponds to computing random walks of ``higher-lengths'', that is to say random walks with many steps. Step 4 will serve to enhance the elements of a column having higher values. This means, in practice, that the most probable transition from node $j$ will become even more probable compared to the other possible transitions from node $j$. The algorithm converges to a matrix invariant under the action of expansion and inflation. The graph associated to such matrix consists of different star-like components; each of them constitutes a community (or cluster) and its central node can be interpreted as the basin of attraction of the community.  
For a given $r>0$, MCl always converges to the same matrix; for this reason it is classified as a parametric and deterministic algorithm. The parameter $r$ tunes the granularity of the clustering, meaning that a small $r$ corresponds to a few big clusters, while a big $r$ returns smaller clusters. In the limit of $r\rightarrow 1$  only one cluster is detected.
In the present study, an analysis at different levels of granularity has been performed in order to find the value of $r$ which better fits with the experimental data. Then we have represented how the average of the number of clusters changes as a function of $r$ (fig. \ref{fig3}). Finally the value $r=1.5$ has been chosen to study in details how the nodes are organized in clusters.

\section{Results}
\begin{figure}[t]
\begin{center}
\epsfig{file=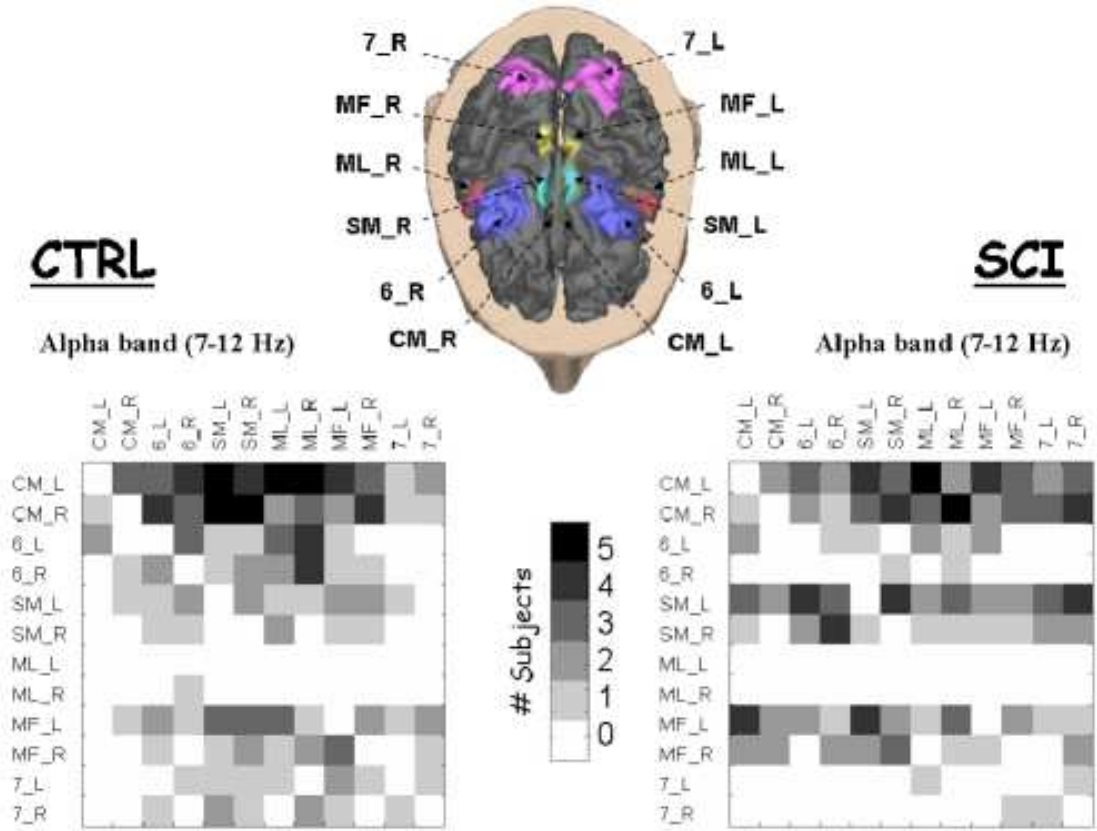,width=0.9\textwidth,angle=0}
\end{center}
\caption{\label{fig1} \emph{Top –} Reconstruction of the head model from magnetic resonance images. The twelve regions of interest (ROIs) are illustrated in colour on the grey cortex and labelled according to previously defined acronyms. \emph{Bottom Left –} Adjaceny matrix for the control group (CTRL) in the Alpha (7-12 Hz) band. The level of grey encodes the number of subjects that hold the functional connection identified by the row $i$ and column $j$. \emph{Bottom Right –} Adjacency matrix for the patients group (SCI) in the Alpha band. Same conventions as above.}
\end{figure}

Fig. \ref{fig1} shows the realistic head model obtained for a representative subject. The twelve ROIs used in the present study are illustrated in colour on the cortex model that is grey coloured. At the bottom side of the fig. \ref{fig1}, we report the adjacency matrices representing the cortical networks estimated, in the Alpha frequency band, from the two analyzed populations during the movement preparation. To  be noticed that such network are directed. Consequently the obtained adjacency matrix is not symmetric. The level of grey within each matrix in figure encodes the number of subjects that hold the functional connection identified by row $i$ and column $j$.

\begin{figure}[t]
\begin{center}
\epsfig{file=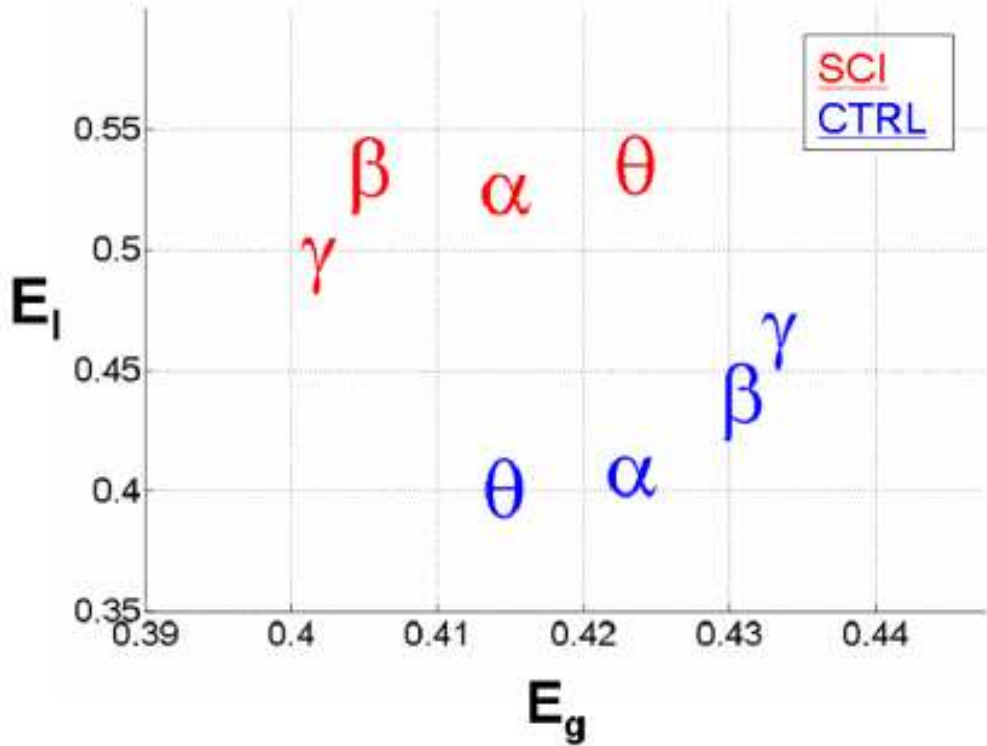,width=0.9\textwidth,angle=0}
\end{center}
\caption{\label{fig2}Scatter plot of the average efficiency indexes obtained from the estimated cortical networks. Global-efficiency is on the $x$-axis, local-efficiency is on the $y$-axis. The Greek symbol encodes the frequency band ($\theta$ Theta, $\alpha$ Alpha, $\beta$ Beta and $\gamma$ Gamma) and it represents the average of the values computed from the control (CTRL, blue-coloured font) and spinal cord injured group (SCI, red-coloured font).}
\end{figure}

As a measure of global and local performance of the network structure we have evaluated the global-efficiency $E_g$ and the local-efficiency $E_l$  indices obtained for each frequency band and for each subject. The average values of $E_g$ and $E_l$ deriving from the healthy group (CTRL) and from the group of patients (SCI) are illustrated for each band in the scatter plot of fig. \ref{fig2}. We have performed an Analysis of Variance (ANOVA) of the obtainedresults. The $E_g$ variable showed no significant differences for the main factors GROUP and BAND. In particular, the ``between'' factor GROUP was found having an F value of 0.83, p=0.392 while the ``within'' factor BAND showed an F value of 0.002 and p=0.99. The ANOVA performed on the $E_l$ variable revealed a strong influence of the between factor GROUP (F=32.67, p=0.00045); while the BAND factor and the interaction between GROUP X BAND was found not significant (F=0.21 and F=0.91 respectively, p values equal to 0.891 and 0.457). Post-hoc tests revealed a significant difference between the two examined experimental groups (SCI, CTRL) in the Alpha and Beta band (p=0.01, 0.03 respectively). In particular, the average local-efficiency of the SCI networks was significantly higher than the CTRL networks for all these bands.

\begin{figure}[t]
\begin{center}
\epsfig{file=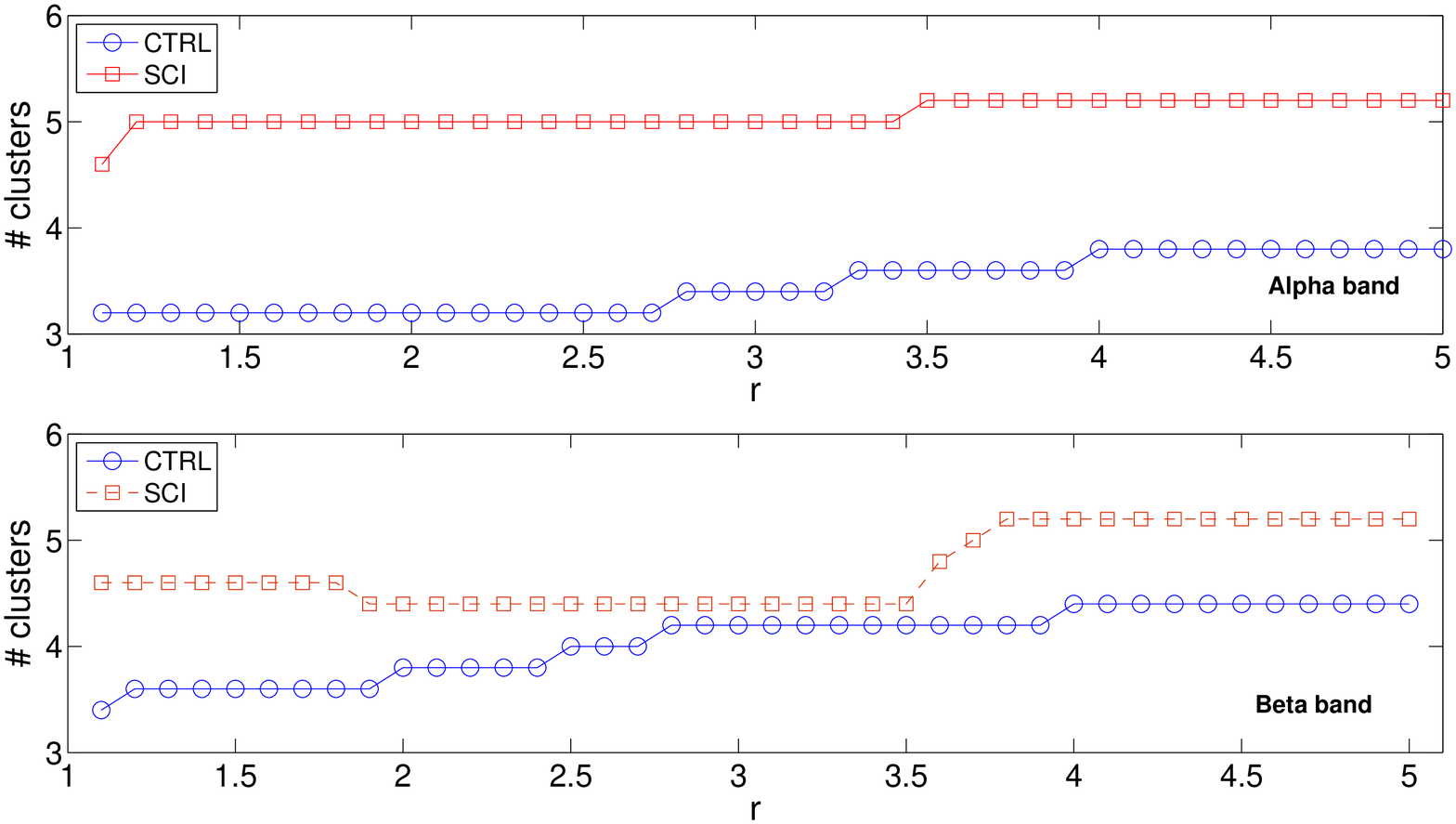,width=1\textwidth,angle=0}
\end{center}
\caption{\label{fig3}Representation of the mean number of clusters for CTRL (blue circles) and SCI (red squares) groups as a function of $r$, granularity parameter of the MCl algorithm, in both the Alpha (top) and Beta band (bottom).}
\end{figure}

The identification of functional clusters within the cortical networks estimated in the control subjects and in the spinal cord injured patients during the movement preparation was addressed through the MCl algorithm (see Methods - Detection of Community Structures (\ref{detection}) ). In fig. \ref{fig3} the average number of clusters detected in the CTRL and SCI networks is reported as a function of the granularity parameter $r$. As it can be observed, for every value of $r$, and for both the Alpha and Beta bands, the average number of clusters is greater for the SCI group than for the CTRL one. One of the main problems with the MCl algorithm is the choice of the value of the granularity parameter to use. Usually good values of $r$ are in the range $]1,3[$ [Vandongen PhD Thesis, 2000; Enright \emph{et al.}, 2002]. For the case under study here, the presence of a plateau in fig. \ref{fig3} indicates that there is a region of values such that the number of clusters does not strongly depends on $r$. We have decided to adopt the granularity $r =1.5$, that is a value in the plateau. For this value of $r$, the average number of clusters in the Alpha band is equal to 3.2 for the cortical networks of the control subjects, while it is equal to 5 for the the spinal cord injured patients. In the Beta band, the average number of cluster is 3.4 for the CTRL networks and 4.6 for the SCI networks, as can be observed at the bottom of the fig. \ref{fig3}.

\begin{figure}[t]
\begin{center}
\epsfig{file=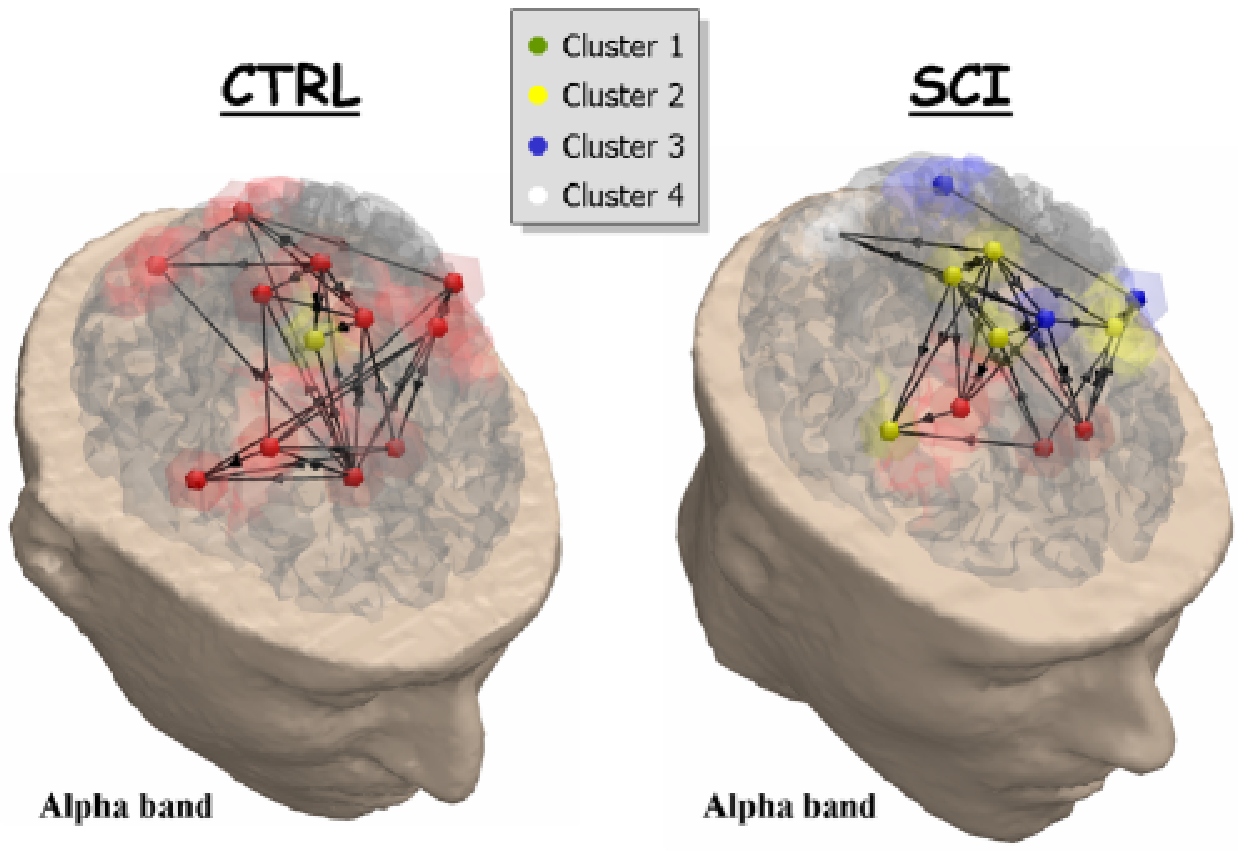,width=0.9\textwidth,angle=0}
\end{center}
\caption{\label{fig4}Graphical representation of the identified clusters of ROIs within the functional networks estimated from a representative control (CTRL) subject and spinal cord injured (SCI) patient during the movement preparation in the Alpha band. The functional network is illustrated as a three-dimensional graph on the realistic cortex model. Spheres located at the barycentre of each ROI represent nodes. Black directed arrows represent edges. The graph partitioning is illustrated through the nodes colouring. Nodes with same colours belong to the same cluster.}
\end{figure}

Fig. \ref{fig4} illustrates the partitioning of the cortical networks estimated in the Alpha band for a representative subject of the control (CTRL) group and for a representative patient of the spinal cord injured (SCI) group. 

Functional networks are represented as three-dimensional graphs on the realistic head model of the experimental subjects. The colour of each node, located in correspondence to each cortical area (ROI), encodes the cluster to which the node belongs. 
The functional clusters detected within the cortical networks in all the subjects and patients participating in the present study are listed in the following tables. Table \ref{tab1} presents the results obtained in the Alpha band while the table \ref{tab2} presents the results obtained in the Beta band.

\begin{table}
\caption{\label{tab1} Cortical network partitioning in the Alpha frequency band.}
\begin{center}
\begin{tabular}{ c | c c c c c c}
\hline
\hline
\multicolumn{1}{ c }{} &
\multicolumn{6}{|c}{\centering{CTRL}} \\
& \hspace{1pt}Cluster 1 \hspace{1pt} & \hspace{1pt} Cluster 2 \hspace{1pt}& \hspace{1pt}Cluster 3 \hspace{1pt} & \hspace{1pt}Cluster 4 \hspace{1pt} & \hspace{1pt} Cluster 5 \hspace{1pt} & \hspace{1pt}Cluster 6\\
\hline
ALFA & CM\_L,CM\_R,6\_L,6\_R, & SM\_R & 0 & 0 & 0 & 0 \tabularnewline & SM\_L,ML\_L,ML\_R,MF\_L,  \tabularnewline & MF\_R,7\_L,7\_R\\ 
\hline
ARGI &  CM\_L,CM\_R,6\_L,6\_R, & SM\_L & SM\_R  & ML\_L & 0 & 0 \tabularnewline & ML\_R,MF\_L,MF\_R,7\_L, \tabularnewline & 7\_R \\
\hline
CIFE & CM\_L,CM\_R,6\_L,6\_R,  & ML\_L  & 7\_L & 0 & 0 & 0 \tabularnewline & SM\_L,SM\_R,ML\_R,MF\_L, \tabularnewline & MF\_R,7\_R\\
\hline
MADA & CM\_L,CM\_R,6\_L,SM\_R, & 6\_R,SM\_L,ML\_R,   & 0 & 0 & 0 & 0 \tabularnewline & ML\_L,7\_L,7\_R & MF\_L,MF\_R\\
\hline
MAMA & CM\_L,CM\_R,6\_L,6\_R,  & ML\_R & 7\_L & 7\_R & 0 & 0 \tabularnewline & SM\_L,SM\_R,ML\_L, \tabularnewline & MF\_L,MF\_R\\
\hline
\hline
 &
\multicolumn{6}{|c}{\centering{SCI}} \\
& \hspace{1pt}Cluster 1 \hspace{1pt} & \hspace{1pt} Cluster 2 \hspace{1pt}& \hspace{1pt}Cluster 3 \hspace{1pt} & \hspace{1pt}Cluster 4 \hspace{1pt} & \hspace{1pt} Cluster 5 \hspace{1pt} & \hspace{1pt}Cluster 6\\
\hline
BARO & CM\_L,CM\_R,6\_R & 6\_L,SM\_R, & SM\_L,ML\_L, & 7\_R & 0 & 0 \tabularnewline & & ML\_R,MF\_L,MF\_R & 7\_L\\
\hline
IRIO & CM\_L,CM\_R,SM\_L, & 6\_L & 6\_R & ML\_L & MF\_R & 7\_R \tabularnewline & SM\_R,ML\_R,MF\_L,7\_L \\
\hline
MASI & CM\_L,CM\_R,6\_L,SM\_L, & 6\_R,MF\_R,7\_R  & 7\_L & 0 & 0 & 0 \tabularnewline & SM\_R,ML\_L,ML\_R,MF\_L \\
\hline
POAL & CM\_L,CM\_R,6\_R,SM\_R & 6\_L, SM\_L, & ML\_L & ML\_R & MF\_L & 7\_R \tabularnewline & & MF\_R,7\_L\\
\hline
TRDA & CM\_L,CM\_R,SM\_L,SM\_R,  & 6\_L & 6\_R & ML\_R & 7\_L & 7\_R   \tabularnewline & ML\_L,MF\_L,MF\_R\\
\hline
\hline
   \end{tabular}
\end{center}
\end{table}

In the Alpha band, the CTRL networks do not present a particularly complicate division into clusters, since for each subject the large part of the cortical areas belong to a unique large community (Cluster 1). In general, the SCI networks are organized in a larger number of clusters more clustered, and two main communities can be observed (Cluster 1 and Cluster 2). The first community is mostly composed by the cingulate motor areas (CM\_L and CM\_R), the supplementary motor areas (SM\_L and SM\_R) and the left primary motor area (MF\_L). The second community is predominantly composed by the left pre-motor areas (6\_L) and the right primary motor area of the foot (MF\_R). The remaining ROIs tend to form isolated groups. 

\begin{table}
\caption{\label{tab2} Cortical network partitioning in the Beta frequency band}
\begin{center}
\begin{tabular}{ c | c c c c c}
\hline
\hline
\multicolumn{1}{ c }{} &
\multicolumn{5}{|c}{\centering{CTRL}} 
\\
& \hspace{1pt}Cluster 1 \hspace{1pt} & \hspace{1pt} Cluster 2 \hspace{1pt}& \hspace{1pt}Cluster 3 \hspace{1pt} & \hspace{1pt}Cluster 4 \hspace{1pt} & \hspace{1pt} Cluster 5 \\
\hline
ALFA & CM\_L,CM\_R,6\_L,6\_R, & 7\_L & 7\_R & 0 & 0 \tabularnewline & SM\_L,SM\_R,ML\_L,ML\_R, \tabularnewline & MF\_L,MF\_R \\
\hline
ARGI &  CM\_L,CM\_R,6\_R,ML\_L, & 6\_L,SM\_L,MF\_R & SM\_R & 7\_R & 0 \tabularnewline & ML\_R,MF\_L,7\_L \\
\hline
CIFE & CM\_L,CM\_R,6\_L,SM\_L, & 6\_R,ML\_R  & 7\_L & 0 & 0 \tabularnewline & SM\_R,ML\_L,MF\_L,\tabularnewline & MF\_R,7\_R \\
\hline
MADA & CM\_L,CM\_R,SM\_L,SM\_R,  & 6\_L,6\_R,MF\_L & ML\_L & 7\_R & 0 \tabularnewline & ML\_R,MF\_R,7\_L\\
\hline
MAMA & CM\_L,6\_R,ML\_L,MF\_R & CM\_R,6\_L,SM\_L, & 7\_L & 7\_R & 0 \tabularnewline & & SM\_R,ML\_R,MF\_L\\
\hline
\hline
 &
\multicolumn{5}{|c}{\centering{SCI}} \\
& \hspace{1pt}Cluster 1 \hspace{1pt} & \hspace{1pt} Cluster 2 \hspace{1pt}& \hspace{1pt}Cluster 3 \hspace{1pt} & \hspace{1pt}Cluster 4 \hspace{1pt} & \hspace{1pt} Cluster 5 \\
\hline
BARO & CM\_L,CM\_R,6\_L,6\_R, & SM\_R,ML\_L & 7\_L & 7\_R & 0 \tabularnewline & SM\_L,ML\_R,MF\_L,MF\_R \\
\hline
IRIO & CM\_L,CM\_R,6\_L,SM\_L, & 6\_R & ML\_R & 7\_L & 7\_R \tabularnewline & SM\_R,ML\_L,MF\_L,MF\_R \\
\hline
MASI & CM\_L,CM\_R,6\_L,SM\_L, & 6\_R & ML\_L & MF\_L & 7\_R \tabularnewline & SM\_R,ML\_R,MF\_R,7\_L \\
\hline
POAL & CM\_L,CM\_R,SM\_L, & 6\_L,ML\_R,MF\_R & 7\_L,7\_R & 6\_R & ML\_L \tabularnewline & SM\_R,MF\_L \\
\hline
TRDA & CM\_L,CM\_R,SM\_L,SM\_R, & 6\_R,ML\_R & 6\_L  & ML\_L & 0 \tabularnewline & MF\_L,MF\_R,7\_L,7\_R \\
\hline
\hline
   \end{tabular}
\end{center}
\end{table}

In the Beta band both the cortical networks of the control and spinal cord injured group tend to get organized in two main modules (Cluster 1 and Cluster 2). In particular, while for both the populations the first cluster is principally composed by the cingulate motor areas CM\_L and CM\_R, the SM\_L and SM\_R and the MF\_L and MF\_R, the second cluster does not present a common set of ROIs across the experimental subjects neither in the CTRL and SCI group.

\section{Discussion}
The evaluation of the estimated cortical networks was addressed by means of a set of measures typical of complex networks analysis [Boccaletti \emph{et al.}, 2006; Micheloyannis \emph{et al.}, 2006; Stam \& Reijneveld, 2007; Hilgetag \emph{et al.}, 2000]. We have first computed global ($E_g$) and local efficiency ($E_l$), two measures that allows characterizing the organization of the functional flows in both the inspected populations [De Vico Fallani \emph{et al.}, 2007b]. The results indicate that spinal cord injuries significantly (p$<$0.05) affect only the local properties of the functional architecture of the cortical network in the movement preparation.  The global property of long-range integration between the ROIs within the network did not differ significantly (p$>$0.05) from the healthy behaviour. The higher average value of local efficiency in the SCI group suggests a larger level of the internal organization and a higher tendency to form modules. In particular, this difference can be observed in the two frequency bands –Alpha (7-12 Hz) and Beta (13-29 Hz)- that are already known for their involvement in electrophysiological phenomena related to the preparation and to the execution of limbs movements [Pfurtscheller \& Lopes da Silva, 1999]. 
Although the efficiency indexes describe the network topology concisely, they are not able to give information about the number of modules and their composition within the network. For this reason, the detection of community structures was addressed by means of the Markov Clustering (MCl) algorithm [Vandongen PhD Thesis, 2000; Enright \emph{et al.}, 2002]. The same method has already been used successfully to detect clusters in sequence similarity networks [Enright \emph{et al.}, 2002] and in configuration space networks deriving from free-energy landscapes [Gfeller \emph{et al.}, 2007]. The obtained results reveal a different average number of clusters for the functional networks of the spinal cord injured patients and the control subjects in both the main spectral contents. 
In particular, in the Alpha band the SCI network presents an average number of modules equal to five, while the CTRL network appears to be divided in three groups. This outcome is in accordance with the significant (p$<$0.05) higher level of local-efficiency found in the functional networks of the SCI patients with respect to the control subjects. A high $E_l$ value reflects a high clustering index $C $and therefore a high density of network communities. The cortical areas of the control subjects do not present a clear partitioning in different modules. They rather appear to belong to a unique community, meaning that they are all involved, in the same way, in the exchange of information during the movement preparation. 
The analysis of the functional communities within the networks obtained for the spinal cord patients revealed a higher tendency to form separate clusters. The pre-motor areas (Brodmann 6\_L and 6\_R), the associative regions (Brodmann 7\_L and 7\_R) and the right primary motor area of the foot (MF\_R) break away from the large module that was found in the networks of the CTRL group. In particular, the area MF\_R and the region 6\_L belong to the same cluster in at least three experimental patients. This result reveals the necessity of the SCI networks to hold a more efficient communication between these frontal pre-motor and primary motor structures, which are already known to be active during the successful execution of a simple movement [Ohara \emph{et al.}, 2001]. 
In the Beta band, the average number of identified clusters in the SCI networks and in the CTRL networks is less different. Moreover, the ROIs that appeared to belong to different clusters in the Alpha band are in this case functionally tied in the same community.
In summary, while in the Alpha band the control group mostly presented a unique large cluster, the spinal cord injured patients mainly exhibited two clusters. These two largest communities are mainly composed by the cingulate motor areas with the supplementary motor areas and by the pre-motor areas with the right primary motor area of the foot. This functional separation is thought to be responsible of the highest level of internal organization in the estimated networks and strengthens the hypothesis of a compensative mechanism due to the partial alteration in the primary motor areas because of the effects of the spinal cord injury [De Vico Fallani \emph{et al.}, 2007a].

\section{acknowledgments}
  The present study was performed with the support of the COST EU project NEUROMATH (BMB 0601), of the Minister for Foreign Affairs, Division for the Scientific and Technologic Development, in the framework of a bilateral project between Italy and China (Tsinghua University). This paper only reflects the authors’ views and funding agencies are not liable for any use that may be made of the information contained herein.

\section{Notes}
In the current work, all the estimated functional networks are treated as unweighted and directed graphs. They all have the same number of connections representing the 25\% (for the community structure analysis) and the 30\% (for the efficiency indexes analysis) most powerful links within the network. These particular values belonged to an interval of thresholds (from 0.1 to 0.5), for which results remained significantly stable [De Vico Fallani \emph{et al.}, 2007a].

\section{References}

\noindent
Achard, S. \& Bullmore, E. [2007] 
``Efficiency and cost of economical brain functional networks,'' 
\emph{PloS Comp Biol} {\bf 3(2)}:e17.

Astolfi, L., Cincotti, F., Mattia, D., Babiloni, C., Carducci, F., Basilisco, A., Rossini, P.M., Salinari, S., Ding, L., Ni, Y., He, B. \& Babiloni, F. [2005] 
``Assessing Cortical Functional Connectivity By Linear Inverse Estimation And Directed Transfer Function: Simulations And Application To Real Data,''
\emph{Clin. Neurophysiol.} {\bf 116(4)}, 920-32.

Astolfi, L., Cincotti, F., Mattia, D., Marciani, M.G., Baccalà, L., De Vico Fallani, F., Salinari, S., Ursino, M., Zavaglia, M., Ding, L., Edgar, J.C., Miller, G.A., He, B. \& Babiloni, F. [2007] 
``A comparison of different cortical connectivity estimators for high resolution EEG recordings,'' 
\emph{Human Brain Mapping} {\bf 28(2)}, 143-57.

Babiloni, F., Cincotti, F., Babiloni, C., Carducci, F., Basilisco, A., Rossini, P.M., Mattia, D., Astolfi, L., Ding, L., Ni, Y., Cheng, K., Christine. K., Sweeney, J. \& He, B. [2005] 
``Estimation of the cortical functional connectivity with the multimodal integration of high resolution EEG and fMRI data by Directed Transfer Function,'' 
\emph{Neuroimage} {\bf 24(1)}, 118-31.

Bartolomei, F., Bosma, I., Klein, M., Baayen, J.C., Reijneveld, J.C., Postma, T.J., Heimans, J.J., van Dijk, B.W., de Munck, J.C., de Jongh, A., Cover, K.S. \& Stam, C.J. [2006] 
``Disturbed functional connectivity in brain tumour patients: evaluation by graph analysis of synchronization matrices,'' 
\emph{Clin. Neurophysiol} {\bf 117}, 2039-2049.

Bassett, D.S., Meyer-Linderberg, A., Achard, S., Duke, T.H. \& Bullmore, E. [2006] 
``Adaptive reconfiguration of fractal small-world human brain functional networks,'' 
\emph{Proc. Natl. Acad. Sci.} {\bf 103}, 19518-19523.

Boccaletti, S., Latora, V., Moreno, Y., Chavez, M. \& Hwang D.U. [2006] 
``Complex networks: structure and dynamics,''
\emph{Physics Reports} {\bf 424}, 175-308.

David, O., Cosmelli, D. \& Friston, K.J. [2004] 
``Evaluation of different measures of functional connectivity using a neural mass model,'' 
\emph{Neuroimage} {\bf 21(2)}, 659-73.

De Vico Fallani, F., Astolfi, L., Cincotti, F., Mattia, D., Marciani, M.G., Salinari, S., Kurths, J., Gao, S., Cichocki, A., Colosimo, A. \& Babiloni, F. [2007] 
``Cortical Functional Connectivity Networks In Normal And Spinal Cord Injured Patients: Evaluation by Graph Analysis,'' 
\emph{Human Brain Mapping} {\bf 28}, 1334-46.

De Vico Fallani, F., Astolfi, L., Cincotti, F., Mattia, D., Tocci, A., Marciani, M.G., Colosimo, A., Salinari, S., Gao, S., Cichocki, A. \& Babiloni, F. [2007b] 
``Extracting Information from Cortical Connectivity Patterns Estimated from High Resolution EEG Recordings: A Theoretical Graph Approach'' \emph{Brain Topogr.} {\bf 19(3)}, 125-136.

Enright, A.J., Van Dongen, S. \& Ouzounis, C.A., [2002] 
``An efficient algorithm for large-scale detection of protein families,'' 
\emph{Nucleic Acids Research} {\bf30}, 1575-1584.

Gfeller, D., De Los Rios, P., Caflisch, A. \& Rao, F. [2007] 
``From the Cover: Complex network analysis of free-energy landscapes,'' 
\emph{Procl. Natl. Acad. Sci.} {\bf 104}, 1817-1822.

Eguiluz, V.M., Chialvo, D.R., Cecchi, G.A., Baliki, M., Apkarian, A.V. [2005] 
``Scale-free brain functional networks,'' 
\emph{Phys. Rev. Lett.}, {\bf 94}, 018102-018106.

Flake, G.W., Lawrence, S., Lee Giles, C. \& Coetzee, F.M. [2002] 
``Self-Organization and Identification of Web Communities,'' 
\emph{IEEE Computer} {\bf 35(3)}, 66–71.

Girvan, M. \& Newman M.E.J. [2002] 
``Community structure in social and biological networks,'' 
\emph{Proc. Natl. Acad. Sci.} {\bf 99(12)}, 7821–7826.

Granger, C.W.J. [1969] 
``Investigating causal relations by econometric models and cross-spectral methods'' 
\emph{Econometrica} {\bf 37}, 424–438.

Grave de Peralta Menendez, R. \& Gonzalez Andino, S.L. [1999] 
``Distributed source models: standard solutions and new developments,'' 
in \emph{Analysis of neurophysiological brain functioning} ed. Uhl, C. (Springer-Verlag, Berlin) pp. 176-201.

Grigorov, M.G. [2005] 
``Global properties of biological networks,'' 
\emph{DDT} {\bf 10}, 365-372.

Guimer\`a, R. \& Amaral L.A.N. [2005] 
``Functional cartography of complex metabolic networks,''
\emph{Nature} {\bf 433}, 895-900.

Harary, F. \& Palmer, E.M. [1973] 
\emph{Graphical enumeration} (Academic Press, New York), p. 124.

Hilgetag, C.C., Burns, G.A.P.C., O'Neill, M.A., Scannell, J.W. \& Young, M.P. [2000] 
``Anatomical connectivity defines the organization of clusters of cortical areas in the macaque monkey and the cat,'' 
\emph{Philos. Trans. R. Soc. Lond. B. Biol. Sci.} {\bf 355}, 91–110.

Horwitz, B. [2003] 
``The elusive concept of brain connectivity,''
\emph{Neuroimage}, {\bf 19}, 466-470.

Kaminski, M., Ding, M., Truccolo, W.A. \& Bressler, S. [2001] 
``Evaluating causal relations in neural systems: Granger causality, directed transfer function and statistical assessment of significance,'' 
\emph{Biol. Cybern.} {\bf 85}, 145-157.

Krause, A.E., Frank, K.A., Mason, D.M., Ulanowicz, R.E. \& Taylor, W.W. [2003] 
``Compartments exposed in food-web structure,'' 
\emph{Nature} {\bf 426}, 282–285.

Lago-Fernandez, L.F., Huerta, R., Corbacho, F. \& Siguenza, J.A. [2000]
``Fast response and temporal coherent oscillations in small-world networks,'' 
\emph{Phys. Rev. Lett.} {\bf 84}, 2758–61.

Latora, V. \& Marchiori, M. [2001] 
``Efficient behaviour of small-world networks,'' 
\emph{Phys. Rev. Lett.} {\bf 87}, 198701.

Latora, V. \& Marchiori, M. [2003] 
``Economic small-world behavior in weighted networks,'' 
\emph{Eur. Phys. J. B.} {\bf 32}, 249–263.

Latora, V. \& Marchiori, M. [2005] 
``Vulnerability and protection of infrastructure networks,'' 
\emph{Phys. Rev. E} {\bf 71}, 015103R.

Latora, V. \& Marchiori, M. [2007] 
``A measure of centrality based on network efficiency,'' 
\emph{New Journal of Physics} {\bf 9}, 188.

Lee, L., Harrison, L.M. \& Mechelli, A. [2003] 
``The functional brain connectivity workshop: report and commentary,'' 
\emph{Neuroimage} {\bf 19}, 457-465.

Lusseau, D. \& Newman, M.E.J. [2004] 
``Identifying the role that animals play in their social networks,'' 
\emph{Proceedings of the Royal Society of London B} {\bf 271}, S477–S481.

Micheloyannis, S., Pachou, E., Stam, C.J., Vourkas, M., Erimaki, S. \& Tsirka, V. [2006] 
``Using graph theoretical analysis of multi channel EEG to evaluate the neural efficiency hypothesis,'' 
\emph{Neuroscience Letters} {\bf 402}, 273–277.

Milgram, S. [1967]
``The Small World Problem,'' 
\emph{Psychology Today} {\bf 1(1)}, 60-67.

Newman, M.E.J. [2003]
``The structure and function of complex networks,'' 
\emph{SIAM Review} {\bf 45}, 167-256.

Ohara, S., Mima, T., Baba, K., Ikeda, A., Kunieda, T., Matsumoto, R., Yamamoto, J., Matsuhashi, M., Nagamine, T., Hirasawa, K., Hon. T., Mihara, T., Hashimoto, N., Salenius, S. \& Shibasaki, H. [2001] 
``Increased synchronization of cortical oscillatory activities between human supplementary motor and primary sensorimotor areas during voluntary movements,'' 
\emph{J, Neurosci.} {\bf 21(23)}, 9377-9386.

Palla, G., Der´enyi, I., Farkas, I. \& Vicsek, T. [2005] 
``Uncovering the overlapping community structure of complex networks in nature and society,'' 
\emph{Nature} {\bf 435}, 814-818.

Pfurtscheller, G. \& Lopes da Silva, F.H. [1999] 
``Event-related EEG/MEG synchronization and desynchronization: basic principles,'' 
\emph{Clin. Neurophysiol.} {\bf 110(11)}, 1842-57.

Pimm, S.L. [1979] 
``The structure of food webs,'' 
\emph{Theoretical Population Biolology} {\bf 16}, 144–158.

Salvador, R., Suckling, J., Coleman, M.R., Pickard, J.D., Menon, D. \& Bullmore, E. [2005] 
``Neurophysiological Architecture of Functional Magnetic Resonance Images of Human Brain,'' 
\emph{Cereb Cortex} {\bf 15(9)}, 1332-42.

Sporns, O., Tononi, G. \& Edelman, G.E. [2000]
``Connectivity and complexity: the relationship between neuroanatomy and brain dynamics,'' 
\emph{Neural Netw.} {\bf 13}, 909–922.

Sporns, O. [2002] ``Graph theory methods for the analysis of neural connectivity patterns,'' 
In R. K\"otter (Ed.), Neuroscience databases, \emph{A practical guide},  171-185.

Sporns, O., Chialvo, D.R., Kaiser, M. \& Hilgetag, C.C. [2004] 
``Organization, development and function of complex brain networks,'' 
\emph{Trends Cogn. Sci.} {\bf 8}, 418-25.

Sporns, O. \& Zwi, J.D. [2004] 
``The small world of the cerebral cortex,'' 
\emph{Neuroinformatics} {\bf 2}, 145-162.

Stam, C.J. [2004] 
``Functional connectivity patterns of human magnetoencephalographic recordings: a 'small-world' network?,'' 
\emph{Neurosci. Lett.} {\bf 355}, 25-8.

Stam, C.J., Jones, B.F., Manshanden, I., van Cappellen van Walsum, A.M., Montez, T., Verbunt, J.P., de Munck J.C., van Dijk B.W., Berendse, H.W. \& Scheltens, P. [2006] 
``Magnetoencephalographic evaluation of resting-state functional connectivity in Alzheimer's disease,'' 
\emph{Neuroimage} {\bf 32(3)}, 1335-44.

Stam, C.J., Jones, B.F., Nolte, G., Breakspear, M. \& Scheltens, P. [2006] 
``Small-World Networks and Functional Connectivity in Alzheimer's Disease,'' 
\emph{Cereb. Cortex}, {\bf 17(1)}, 92-9.

Stam, C.J. \& Reijneveld, J.C. [2007]
``Graph theoretical analysis of complex networks in the brain,'' 
\emph{Nonlinear Biomed. Phys.} {\bf 1}, 3 epub.

Strogatz, S.H. [2001] 
``Exploring complex networks,'' 
\emph{Nature} {\bf 410}, 268-76.

Tononi, G., Sporns, O. \& Edelman, G.M. [1994] 
``A measure for brain complexity: relating functional segregation and integration in the nervous system,'' 
\emph{Proc. Natl. Acad. Sci.} {\bf 91}, 5033–5037.

Van Dongen, S., [2000]
\emph{Graph Clustering by Flow Simulation}, 
PhD thesis, University of Utrecht.

Watts, D.J. \& Strogatz, S.H. [1998] 
``Collective dynamics of 'small-world' networks,'' 
\emph{Nature} {\bf 393}, 440-442.

\end{document}